\documentclass[rnote]{aa}
\usepackage{graphicx}
\usepackage{natbib} 
\bibpunct{(}{)}{;}{a}{}{,} 
\usepackage{txfonts}
\begin{document}
\newcommand{\MPIA}{1} 
\newcommand{\Ms}{{2MASS\,J0756}}
\newcommand{\Mn}{{2MASS\,J0939}}
\newcommand{\Mt}{{2MASS\,J1036}}
\newcommand{\SD}{{SDSS\,J2047}}
\newcommand{\IPMS}{{SIMP\,J0136}}
%
   \title{Binarity at the L/T brown dwarf transition}

   \subtitle{Adaptive optics search for  companions.\thanks{Based on observations collected at the European Observatory, Paranal, Chile,
             under programmes 78.C-0754 and 79.C-0635. 
             }
   }

   \author{
       B. Goldman\inst{\MPIA}
          \and
       H. Bouy\inst{2,3}\thanks{Marie Curie Outgoing International Fellow MOIF-CT-2005-8389.}
          \and
        M.~R. Zapatero Osorio\inst{2}
          \and
       M.~B. Stumpf\inst{\MPIA}
          \and
      W. Brandner\inst{\MPIA}
          \and
       T. Henning\inst{\MPIA}
          }

   \offprints{B.~Goldman, {\tt go{\,\hspace{-1pt}}ld\hspace{-1pt}\,man{\,}@mp{}ia.de}}

   \institute{
              Max Planck Institute for Astronomy, K\"onigstuhl~17, D--69117 Heidelberg, Germany
         \and 
              Instituto de Astrof{\'\i}sica de Canarias, C/ V\'\i a Lact\'ea S/N, E-38205 La~Laguna, Tenerife, Spain
         \and 
           Astronomy Department, 601 Campbell Hall, UC Berkeley, Berkeley CA-94720-3411, U.S.A 
       }

   \date{Received 15 July 2008 / Accepted 17 August 2008}

   \abstract
   {Current atmospheric models cannot reproduce some of the characteristics of the transition between the L dwarfs with cloudy atmospheres and the T dwarfs with dust-depleted photospheres. 
   It has been proposed that a majority of the L/T transition brown dwarfs could actually be a combinaison of a cloudy L dwarf and a clear T dwarf. Indeed
   binarity seems to occur more  frequently among L/T transition brown dwarfs.
   }
   {We aim to refine the statistical significance of the seemingly higher frequency of binaries. Co-eval binaries would also be interesting test-beds for evolutionary models. } 
   {We obtained high-resolution imaging for six mid-L to late-T dwarfs, with photometric distances between 8 and 33\,pc, using the adaptive optics systems NACO at the VLT, and the Lick system, both with the laser guide star. 
  }
   {We resolve none of our targets. Combining our data with published results, we obtain a frequency of resolved L/T transition brown dwarfs of $31^{+21}_{-15}$\%, compared to $21^{+10}_{-7}$\% and $14^{+14}_{-7}$\% for mid-L and T dwarfs (90\% of confidence level).
   These fractions do not significantly support, nor contradict, the hypothesis of a larger binary fraction in the L/T transition.
   None of our targets has companions with effective temperatures as low as 360--1000\,K at separations larger than 0\farcs5.}
  {}
  
  
   \maketitle
%

\section{Introduction}

  L/T transition brown dwarfs are a class of sub-stellar objects comprising the latest L-type
  and the earliest T-type dwarfs {\citep{Legge00}, with spectral types between roughly L8 and T4 \citep{Loope08}}.  
  {Historically, the warmest members of the group, the late L dwarfs, were soon identified through their very red near-infrared colours \citep{Marti99,Kirkp00},
   but the T-type members of the transition kept escaping detection even after 
   the cooler T dwarfs were identified through their 
   strong molecular absorption (``methane dwarfs''). 
   Indeed the neutral near-infrared colours of cooler L/T transition brown dwarfs concealed them to near-infrared surveys such as 2MASS, while the optical/near-infrared survey DENIS was too shallow to provide a large-enough volume \citep{PhanB08}. Eventually \citet{Legge00} identified the first T~members of this group in the SDSS EDR data.
  Their near-infrared spectra} exhibit both CO and CH$_4$ absorption and 
  they have $J-K$ colours {filling the gap} between those of the red late-type L dwarfs and the blue mid-T dwarfs ({i.e. $0\lesssim J-K\lesssim$2.0}).  

 \begin{table*}
  \caption[]{Summary of targets and references.
  Tip-tilt star magnitude and separation to the target based on USNO-A2.0 \citep{Monet98}. 
  All distances are photometric, based on the relations of \citet{Knapp04} between the spectral type, and $M_J$, and $M_K$ (except for \SD), assuming non-binarity. 
  }
  \label{targets}
  \begin{tabular}{lccccccc}
    \hline
    \hline
    \noalign{\smallskip}
    NACO Targets                                 & tip/tilt($R$)               & Sp.T. $^1$&           $J\;^2$ (mag)       &                      $H\;^2$ (mag)          & dist. (pc) & $\mu_\alpha$ (mas/yr) & $\mu_\delta$ (mas/yr) \\
    \noalign{\smallskip}
    \hline
    \noalign{\smallskip}
    \object{2MASSI J0756252+124456}$^3$        & $16.3, 37\arcsec$ & L6$^3$ & $16.66\pm 0.14$ & $15.76\pm 0.15$ & 38 & $-3\pm13$ & $-93\pm43$ \\ 
    \object{2MASS J09393548$-$2448279}$^4$ & $12.2, 30\arcsec$ & T8 & $15.98\pm 0.11$ & $15.80\pm 0.15$ & 9  & 490$^4$ & $-1040$$^4$ \\ 
    \object{2MASSW J1036530$-$344138}$^5$ & $17.6, 17\arcsec$& L6$^5$ &  $15.62\pm 0.05$ & $14.45\pm 0.04$ & 22  & $-123\pm20$ & $-426\pm20$ \\\ 
    \object{SDSS J204749.61$-$071818.3}$^6$ & $13.8, 12\arcsec$ & T0: & $16.47\pm 0.20$ & $15.91\pm 0.20$ & 32 & $-3\pm20$ & $-274\pm20$ \\ 
    \noalign{\smallskip}
    \hline
    \noalign{\smallskip}
    Lick Targets                                   & tip/tilt($R$)               & Sp.T. &           $J$ (mag)      &                      $Ks$ (mag)           & dist. (pc) & $\mu_\alpha$ (mas/yr) & $\mu_\delta$ (mas/yr) \\
    \noalign{\smallskip}
    \hline
    \noalign{\smallskip}
    \object{2MASS J16262034+3925190}$^7$        & $14.5, 35\arcsec$ & dL4$^8$ & $14.44\pm 0.03$ & $14.47\pm 0.07$ & 37$^9$  & $+286^7$ & $-1237^7$\\ 
    \object{{SIMP} J013656.5+093347.3}$^{10}$           &                                  & T2.5$^{10}$ & $13.46\pm 0.03$ & $12.56\pm 0.03$ & 7.6$^{10}$  & $-4^{10}$ & $+1241^{10}$\\  
    \noalign{\smallskip}
    \hline
  \end{tabular}
  \begin{tabular}{l}
$^1$  Spectral types from \citet{Burga06cl} except for \Ms,  \Mt\ , 2MASS~J1626 and \IPMS. \\
$^2$  Target photometry from 2MASS \citep{Skrut06}. \\
$^3$ \citet{Kirkp00}. \\ 
$^4$ \citet{Tinne05}. \\ 
$^5$ \citet{Gizis02}. \\
$^6$ \citet{Knapp04}. \\ 
$^7$ \citet{Burga04sd}. \\
$^8$ \citet{Burga07sL} \\ 
$^9$ The distance is based on the $K$ magnitude only, as the parallax measurement of sdL7 subdwarf revealed a $M_K$ similar to the \\ \hskip 2mm solar-metallicity L7 dwarfs, while the subdwarf is more than 1~magnitude brighter in the $J$ band \citep{Burga08sd}. \\ 
$^{10}$  \citet{Artig06}. They report a distance of $6.4\pm 0.4$\,pc. \\ 
  \end{tabular}
\end{table*}
  
  The rapid transition between the cloudy atmospheres of the L dwarfs and the dust-depleted atmospheres of the T~dwarfs, at nearly constant effective temperature, was quickly recognised as a challenge for our theoretical understanding \citep{Kirkp00,Golim04}. A brightening in the $J$ band is even observed for early T~dwarfs \citep{Dahn02,Tinne03,Vrba04,Loope08}. Other pecularities include the resurgence of FeH bands and the restrengthening of 1.2-$\mu$m K\,{\sc i} lines. Several unsuccessful attempts have been made to explain these peculiarities through a global thinning of the dust cover \citep{Tsuji05}, a variation in the precipitation efficiency \citep{Knapp04}, and a better description of the condensate cloud structure \citep{Coope03,Burro06}. This led \citet{Burga02lt} to propose that a dynamical cloud fragmentation process takes place in the atmospheres of these objects. So far a characteristic weather-like behaviour that this fragmentation could produce has not been clearly identified \citep[see e.g.][]{Goldm08}.
  
  Recent high-resolution imaging surveys have revealed a possibly higher binary frequency among L/T transition objects, compared to other brown dwarfs \citep{Burga06bi,Liu06}. 
  The factor-of-two excess reported by \citet{Burga06bi}, given the limited sample size (19\,dwarfs over the L7 to T3.5 spectral types in their compilation), would not rule out a constant binary fraction at the 90\% confidence level \citep{Burga07lt}.
  This latter article demonstrates that such a higher binary fraction is expected where the $M_{bol}/$spectral type relationship flattens, under a wide range of (reasonable) formation histories and mass functions, particularly in magnitude-limited samples.
 \citet{Burga06bi} and \citet{Liu06} formulated the hypothesis that an higher binary rate among the transition brown dwarfs would explain some of the above characteristics, with \citet{Liu06} speculating that single $\approx$ T2--T4.5 objects may be rare.
{ Their results, as well as the binary 2MASS\,J14044941$-$3159329 found by \citet{Loope08}, also demonstrated that the $J$-band brightening across the L/T transition is real and not an effect of selection bias, as it affects binaries which presumably have the same age and metallicity.}

  Co-eval binaries also provide interesting tests to evolutionary models, as the usually quite uncertain parameters such as age and metallicity, are now fixed for the pair of objects \citep{Bouy08bi,Loope08}. Finally, the binary fraction is an important prediction of brown dwarf formation scenarios, and, with other parameters, will eventually allow to discriminate between the various scenarios \citep{Biller05,Umbre05,Cabal07k1}.
  {In particular, \citet{Allen07bs} has analysed several high-resoution imaging surveys using a Bayesian algorithm. He substantiated the high-mass ratio distribution, and argued that the observed fraction of binaries with separation larger than 2.5\,A.U. is too large compared to that predicted by \citet{Bate05}.}
 
  Brown dwarfs are too faint in the optical to be good natural guide stars for adaptive optics high-resolution imaging, and even near-infrared sensors are not sensitive enough for most of the brown dwarfs, but the closest ones. Laser guide star systems have recently extended the fraction of the sky observable with adaptive optics, although their operation does require a nearby star for tip-tilt corrections.

Here we report on an on-going search for binarity among L/T transition objects.
In Section\,\ref{obs}, we present our observations, target list and reduction, 
and in Section\,\ref{results} our results.
Finally in Section\,\ref{discussion} we interpret our (non) detections. 


\section{Observations and data reduction} \label{obs}

\subsection{Target selection}

We searched through the compendium housed at {\tt DwarfArchives.org} and chose our targets according to the following criteria:
\begin{enumerate}
\item Availability and characteristics of the tip-tilt star. Laser-guide-star systems require a natural guide star to correct for the tip-tilt motions. Better Strehl ratios are obtained for brighter guide stars, that are closer to the target. 
We searched the USNO-A2.0 catalogue \citep{Monet98} at CDS, Strasbourg, for appropriate tip-tilt stars. 
For the ESO laser system PARSEC, we selected targets later than L4, with $R\leq 17.6$\,mag, $d<40\arcsec$ guide stars. 
For the Lick laser system, we selected targets with $R\leq 17.0$\,mag, $d<50\arcsec$ guide stars.
\item The targets should not have been observed with high-resolution imagers at ESO and on board HST, and no such observations should have been published. 
\item Spectral type: we gave precedence to dwarfs with spectral type close to T0.
\item Brightness: for a given spectral type we preferred brighter targets as they are likely to be closer to us and allow detections of closer binaries.
\item Observability: The Lick run was performed in visitor mode and we selected target with the best observability at the time of the observations. The ESO observations were performed in service mode and we only required that the targets be South of $\delta = +20\deg$.
\end {enumerate}
The targets are described in Table\,\ref{targets}. 

\begin{figure*}]t]
\includegraphics[width=\textwidth]{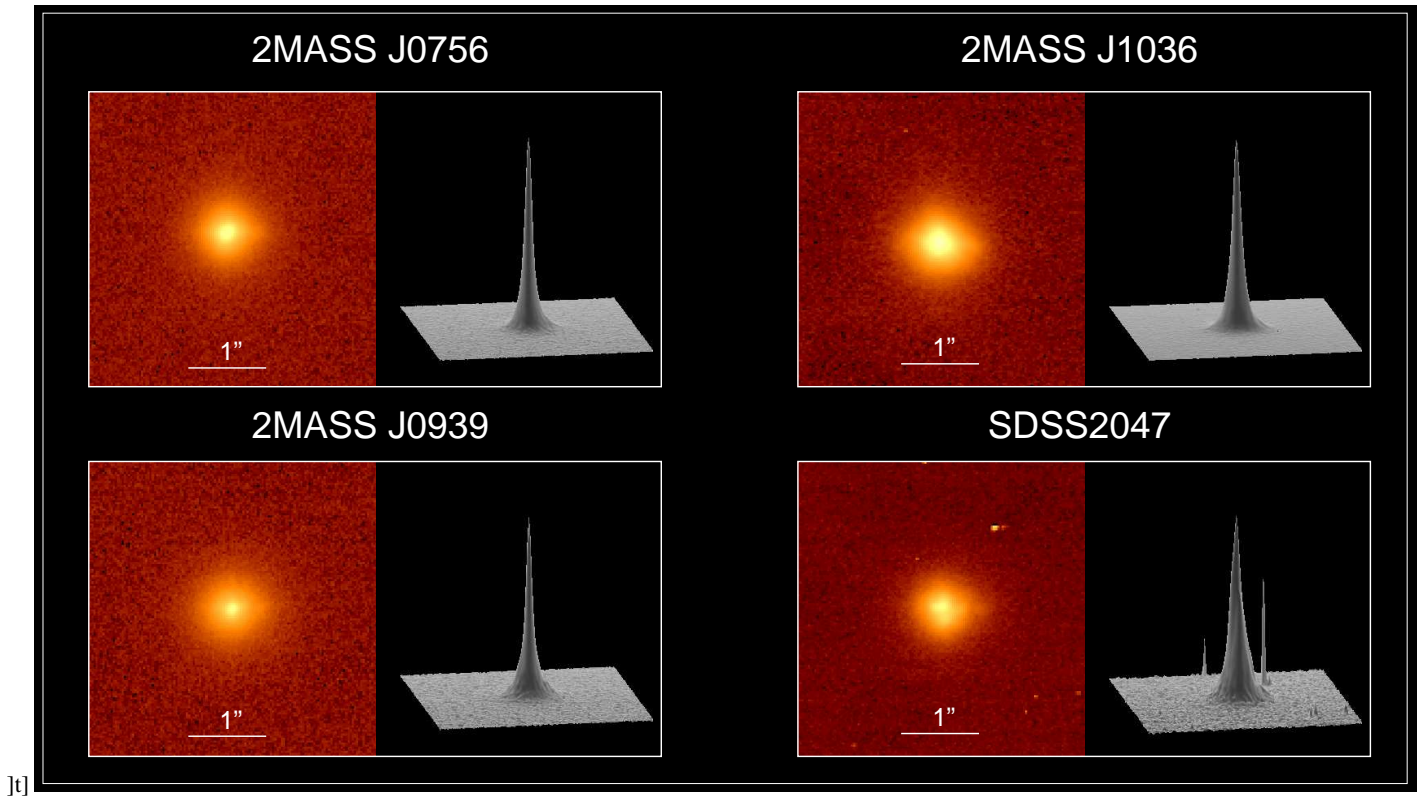}
   \caption{NACO $H$-band  images.
           }
      \label{NACO-img}
\end{figure*}

\subsection{NACO/PARSEC observations}

We observed all targets in $J$ and $H$ bands to strengthen any detection and to derive a rough spectral type for the potential companion, based on photometry.
We used the S27 camera with a 27-mas pixel size and $28\arcsec\times28\arcsec$ field of view. This camera maximises the field of view, while offering a good sampling of the point-spread function.
In order to remove, or at least characterise, instrumental effects, we observed each target at two instrumental angles, $0\deg$ and $30\deg$. For each angle, we used a four-position dithering pattern, with one and two exposures/position, respectively in the $J$ and $H$ band, for a total integration time per angle of 4\,min and 9.6\,min.
The observations were performed in service mode.
The nights were clear (except on April~2) but the seeing was often highly variable. 
However, the seeing of the final, stacked images barely varies, within a few percent, with the exception of \Mn, as well as \SD\ in the $H$ band. 

We have used the reduced images produced by the NACO pipeline, version 3.6.0 (see Fig.\,\ref{NACO-img}).
The pipeline determines the astrometry of the image (WCS), with a typical accuracy of 1\% for the pixel scale and $0.4\deg$ for the orientation \citep{Eggen07}.

\subsection{Lick observatory adaptive optics observations} 

We used the Lick Observatory 3-m Shane telescope and its adaptive optics system \citep{Bauma99,Lloyd00} to observe the L4 subdwarf 2MASS~J1626, 
as well as \IPMS, one of the closest T~dwarf.
With $R\approx$20~mag, the target was too faint for the wavefront sensor to close the loop, so that we used the laser guide star. 
We acquired three dithered images of 5~min each in th $Ks$-band. Conditions were average for the site, with a seeing of $\approx$1\farcs5 and a clear sky. 

\begin{figure}
\includegraphics[width=.5\textwidth]{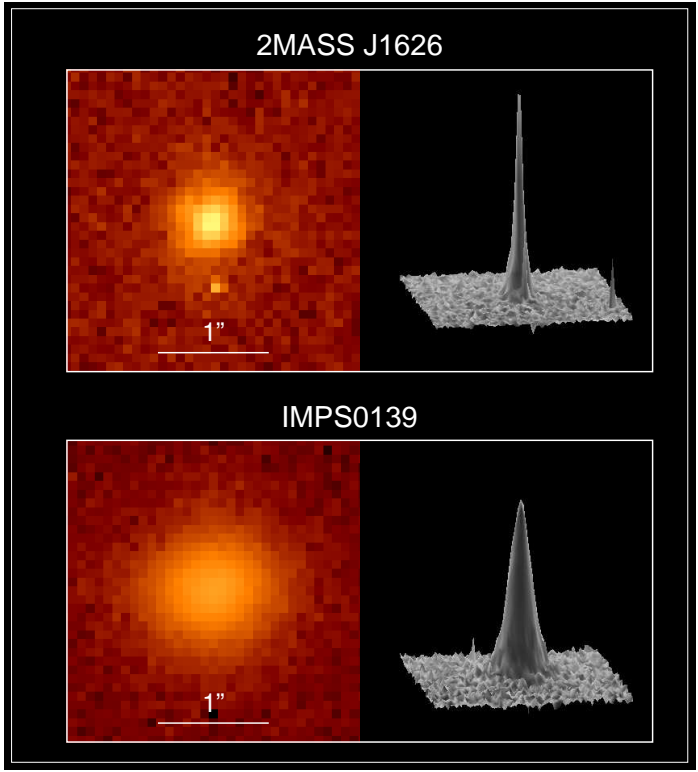}
   \caption{Lick $Ks$-band images.
     The spike South of 2MASS~J1626 is much narrower than the point-spread function and cannot be stellar.
           }
      \label{Lick-img}
\end{figure}

The images were processed using standard procedures using the Eclipse data reduction package \citep{Devil97}, including dark subtraction, flat-fielding, sky extimation and co-addition of the individual frames. The images are shown in Fig.\,\ref{Lick-img}.

%
%

\begin{table}[b]
  \caption[]{Observing log. Site seeing from the DIMM.} 
  \label{Obs}
  \begin{tabular}{lr@{ }lcc}
    \hline
    \hline
    \noalign{\smallskip}
    NACO Target &   \multicolumn{2}{c}{Start UTC}  & seeing & airmass \\
    \noalign{\smallskip}
    \hline
    \noalign{\smallskip}
   \Ms   & 24/03/2007 & 00:39   & 0.75--1.20\arcsec &  1.25--1.29 \\ 
   \Mn   & 20/03/2007 & 04:08   & 1.00--1.20\arcsec & 1.07--1.17 \\ 
   \Mn   &   2/04/2007 & 04:41   & 0.71--1.12\arcsec & 1.32--1.42 \\  
   \Mt    &  21/02/2007 & 04:36  & 0.50--0.65\arcsec &  1.02--1.05 \\ 
   \SD   & 12/09/2007 & 03:15   & 0.65--1.13\arcsec & 1.09--1.17 \\ 
    \noalign{\smallskip}
    \hline
     \noalign{\smallskip}
    Lick Target &   \multicolumn{2}{c}{Start UTC}  & seeing & airmass \\
    \noalign{\smallskip}
    \hline
    \noalign{\smallskip}
    2MASS~J1626 & 3/04/2007 & 10:24   & 1.5\arcsec & 1.03-1.07 \\
    \IPMS                  & 1/09/2007&  09:56   & 1.2\arcsec & 1.15--1.17\\
    \noalign{\smallskip}
    \hline
  \end{tabular}
\end{table}


\subsection{New proper motions}

Three of our NACO targets have objects in the field of view that have multi-epoch detections (see Section\,\ref{widebin}) and which can be used to constrain the image astrometry and to derive the proper motions of our targets.
Proper motions are crucial to confirm the association with potential companions.
Because there are fluctuations in the NACO pixel scale and orientation of up to 1\% and $0.4\deg$ respectively, an accurate astrometry would require two objects with known proper motion. 
As we only have one, the above uncertainties translate into a 120-mas (200-mas) and 84-mas (140-mas) uncertainties for a field object located at 12\arcsec\ (20\arcsec\ respectively) away from the target.
Depending on the time baseline, those uncertainties may be greater than the proper motion uncertainties of the field object.

\subsubsection{\Ms}

\Ms\ has been observed twice by SDSS \citep{Adelm08} in $2005.00\pm0.05$ and once by 2MASS \citep{Skrut06} in 1997.83. We derive the proper motion based on the published positions (see Table\,\ref{targets}).

Using that proper motion, we measure that the pixel scale is within 2\% of the nominal value of 27.15\,mas, while the orientation is within $(0.57\pm0.15)\deg$ of the nominal value.

\subsubsection{\Mt}

\Mt\ has a nearby $J=15.15$-mag field star, whose (negligible) proper motion we derive from USNO and 2MASS catalogues. 
Adding this proper motion to the positional shift in our images compared to the 2MASS position, we obtain a proper motion of $\mu_\alpha=(-0.12\pm0.02)$\arcsec/yr, $\mu_\delta=(-0.43\pm0.02)$\arcsec/yr for the brown dwarf.
We use the nominal pixel scale and orientation, but the large baseline of 7.9\,years reduces the impact of those uncertainties (included in the quoted errors).

\Mt\ is also detected in 2MASS and DENIS images, separated by 2.1\,years. 
We cross-matched all DENIS detections of Strip\,5346 within 5\,min of the target, to the 2MASS catalogue (138\,objects). We find an average offset of \mbox{$\Delta\alpha=+159\pm16$\,mas} and $\Delta\delta=-380\pm20$\,mas (stat., $1\sigma$), independent of brightness, nearly constant over the field except for a trend in $\Delta\delta$ with right ascension. An accurate determination of the relative DENIS-2MASS astrometry would require a more detailed treatment and is beyond the scope of this paper. Applying these corrections to the DENIS astrometry improves the proper motion fit of the field star. 
However the resulting proper motion of \Mt\  is found to be incompatible with the above value, calculated over a longer baseline, which we favour.

\subsubsection{\SD}

\SD\  has a nearby $J=12.56$-mag field star, with negligible proper motion. 
As above we use the relative SDSS and NACO positions to derive the proper motion of \mbox{$\mu_\alpha=(-0.00\pm0.02)$\arcsec/yr}, \mbox{$\mu_\delta=(-0.27\pm0.02)$\arcsec/yr}, with a 7.0-yr baseline. 

\section{Results} \label{results}

\subsection{Limit of sensitivity of the adaptive optics observations}

\begin{figure}[b]
\includegraphics[width=.5\textwidth]{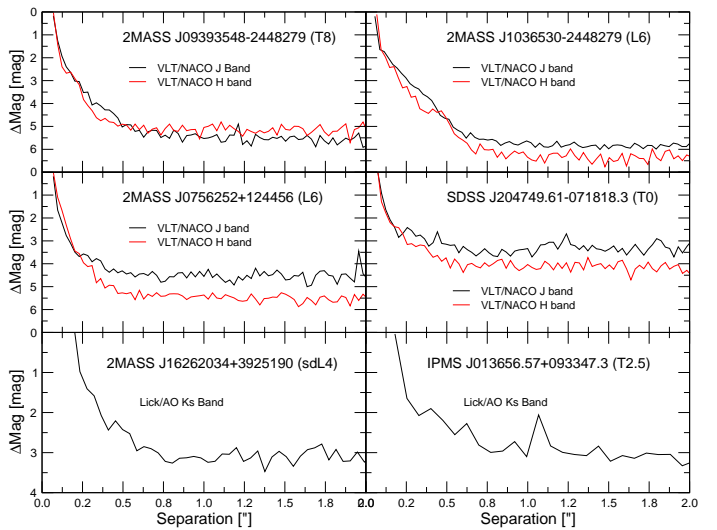}
   \caption{Limits of sensitivity $\Delta$Mag vs separation of the adaptive optics observations performed with VLT/NACO ($J$ and $H$ bands) and Lick/AO ($Ks$ band).
           }
      \label{sensitivity_ao}
\end{figure}

We have computed the limit of sensitivity of our observations from the 3-$\sigma$ noise on the radial profile of the PSF of the different images. Figure \ref{sensitivity_ao} shows the results. NACO $H$-band observations are typically sensitive to $\Delta H=5$--6~mag at 0\farcs5, and $\Delta H\approx 4$~mag at 0\farcs25. The Lick/AO observation is sensitive to $\Delta Ks=$3~mag at 0\farcs5, and $\Delta Ks=$1.5~mag at 0\farcs25.

With these 3-$\sigma$ sensitivities, we calculate the absolute $J$- and $H$-band limiting magnitudes at 0\farcs5, assuming the distances listed in Table\,\ref {targets}.
We translate these absolute magnitudes into effective temperature for the most sensitive filter, usually the $J$~band for the NACO data, assuming the COND 5-Gyr tracks of \citet{Allar01} (see Table.\,\ref{limits}).
We note that, although the Strehl ratio in the $J$ band is {worst than in} the $H$ band, the sensitivity is equal or better in the $J$ band than in the $H$ band thanks to the smaller Airy disk and the larger $J$ flux for T~dwarfs and cooler dwarfs.

\begin{table}
  \caption[]{Sensitivities to ultra-cool companions at 0\farcs25/0\farcs5 (3 $\sigma$).
   Lower limits on the effective temperature based on \citet{Allar01}.}
  \label{limits}
  \begin{tabular}{lcccc}
    \hline
    \hline
    \noalign{\smallskip}
    Target & $J$ &  $H$ & $M_J$ or $M_K$ & $T_{\rm eff}$ (K) \\
    \noalign{\smallskip}
    \hline
    \noalign{\smallskip}
   \Ms   & 20.3/21.0 & 19.3/21.0 & 17.4/18.1 & 700/610 \\ 
   \Mn   & 19.5/21.0 & 19.3/20.8 & 19.7/21.2 & 450/360 \\ 
   \Mt    & 18.6/20.2 & 17.5/19.1 & 16.9/18.5 & 760/710 \\ 
   \SD   & 19.0/19.7 & 18.4/19.8 & 16.5/17.2 & 800/730 \\ 
    \noalign{\smallskip}
    \hline
    \noalign{\smallskip}
    2MASS~J1626 &  \multicolumn{2}{c}{$Ks=15.6/16.9$}  & $14.4/15.7$ & 1300/1000 \\ 
    \IPMS &                   \multicolumn{2}{c}{$Ks=14.2/14.9$}  &  $14.9/15.6$ & 1130/1000 \\
    \noalign{\smallskip}
    \hline
  \end{tabular}
\end{table}

\subsection{Search for close binaries}

We resolve no close companion in any of our images.

For the nearly-equal-brightness binaries that could populate the L/T transition ($\Delta m<2$), we are sensitive to separations down to about 0\farcs1 for the NACO data, and 0\farcs2 for \IPMS, corresponding to 1--40\,AU.
If the targets were equal-brightness binaries (unresolved in our data), they would be 41\% further away and the minimum physical separation probed would increase accordingly. 

We update the compilation of high-resolution-imaging results of \citet{Burga07lt}, to which we add the two T2.5 targets observed by \citet{Loope08} (one is resolved) and ours.
Over the L8 to T2.5 spectral types, we find a binary fraction of $31^{+21}_{-15}$\%, compared to $21^{+10}_{-7}$\% and $14^{+14}_{-7}$\% for mid-L and T dwarfs (90\% of confidence level).
The statistical uncertainties are calculated using the formula of \citet{Burga03bi}.
A simple $\chi^2$ analysis of the different binarity fractions returns a 50\% chance result for a 20\% constant binary fraction over the L0 to T8 spectral range.

\begin{figure}[h]
\begin{center}
\includegraphics[width=.5\textwidth]{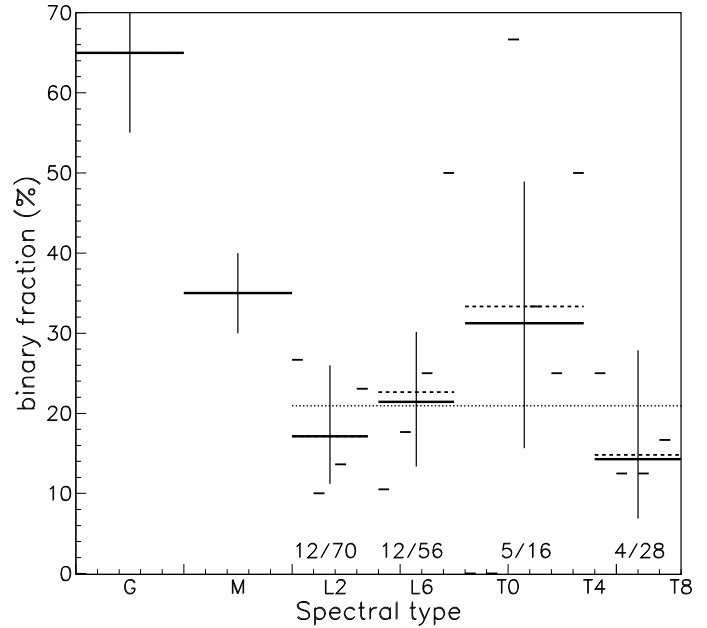}
   \caption{Close binary fraction vs. spectral type.
     The short thick lines show the new binary fraction for each L and T spectral type.
     The longer dashed lines indicate the published fraction, averaged over four groups of subtypes \citep{Burga07lt}, while the thick lines give the updated average fraction including our results. 
     The bottom numbers indicate the numbers of resolved binaries and total numbers of observed brown dwarfs.
     The 90\% confidence levels for statistical noise are reported as vertical thin lines.
     We include the results of the \citet{Duque91}~survey for G~stars and \citet{Reid97}. We point out again the different survey sensitivities, in particular \citet{Duque91} is based on the radial-velocity method.
           }
      \label{binfrac}
\end{center}
\end{figure}

As mentioned in \citet{Burga07lt} {and \citet{Allen07bs}}, the sensitivity of these surveys, both in terms of depth and spatial resolution, varies greatly. Depth is usually not a critical factor for the nearly-equal-brightness binaries we are concerned with, but the physical separations probed, which depend on the target distance and instrumental performance, are crucial.
The binarity fractions of Fig.\ref{binfrac} are obviously lower limits, as very close binaries will not be resolved.
{\citet{Allen07bs} estimates to 3--4\% the fraction of ultra-cool dwarfs with a tight companion undetected in the imaging surveys he analyses.}
If we limit ourselves to a comparison of the binarity fractions with respect to spectral types, we need still to address the biases introduced by the incomplete brown dwarf detection and by the target follow-up selections. 
For instance, our samples are primarily magnitude-limited; L/T transition brown dwarfs are mostly detected in an optical survey (SDSS) limited over a quarter of the sky; the smaller number of L/T transition brown dwarfs may systematically lead to the study of more distant targets than, e.g., the (also fainter) later T~dwarfs.
We defer to the presentation of our full sample the evaluation of these statistical biases and survey heterogeneity. 

\subsection{Wide binaries} \label{widebin}

We check if the field stars in the NACO field of view of 28\arcsec\ could be associated with the target. 
\begin{itemize}
\item \Ms\ has an object with $J_{\rm field}-J_{\rm BD}=\Delta J=-0.1$\,mag and \mbox{$\Delta H=0.6$\,-mag}, 19\arcsec\ away in the field, but relative proper motions based on 2MASS \citep{Skrut06}, SDSS--DR6 \citep{Adelm08} and our data clearly reveal a background object.
\item \Mn\ has two faint objects with nearly identical colours \mbox{$J-H\approx1.1$\,mag}, $J=19.4$\,mag and $J=20.0$\,mag, 12 and 14\arcsec\ away respectively. There are no proper motion information available.
They could be L$5\pm 2$ dwarfs at distances larger than 100\,pc, or extragalactic sources.
\item \Mt\ has a source with $\Delta J=-0.3$\,mag and $\Delta H=0.2$\,mag, 13.2\arcsec\ away. The separation in the 2MASS catalogue is 16.6\arcsec, significantly larger than in ours. 
It is therefore a background source.
\item \SD\ has an object with $\Delta J=-0.3$\,mag and $\Delta H=0.2$\,mag, 13.2\arcsec\ away. The separation in the 2MASS catalogue is 12.1\arcsec, significantly smaller than in ours. 
It is therefore a background object.
\end{itemize}
%

None of the field stars are resolved. Because the field stars are off-axis, {their image quality differs from that of the program star. In the case of \SD's field star, its full width at half maximum is much larger than that of \SD\ (by $\approx 50\%$).
For the three other targets, the full width at half maximum of the field star is similar or even better than that of the on-axis target.}

\section{Conclusion} \label{discussion}


  We have obtained high-resolution imaging of six brown dwarfs using Laser guide star systems at the VLT and Lick observatories. No targets are resolved in our data set. 
  A slightly larger fraction of L/T transition brown dwarfs are binaries, compared to earlier L and later T dwarfs, using data from the literature and adding our non-detections. The excess is however statistically not significant.
  
  Observations of a larger sample, particularly of L/T transition brown dwarfs, is required. (Nearly) whole-sky multi-band optical surveys, such as PanSTARRS1, and near-infrared proper motion surveys, such as UKIDSS and VHS, using 2MASS as the first epoch, will offer the opportunity to find additional targets close enough to provide a good physical-separation sensitivity. 
    
\begin{acknowledgements}

       B.~Goldman thanks P.\,Moeller and the ESO staff for its support and for conducting the observations.
H. Bouy acknowledges the funding from the European Commission's Sixth Framework Program as a Marie Curie Outgoing International Fellow (MOIF-CT-2005-8389).
       This Research has made use of the M, L, and T dwarf compendium housed at DwarfArchives.org and maintained by C. Gelino, D. Kirkpatrick, and A. Burgasser, 
       and of the {\sc Simbad} database, operated at C.D.S., Strasbourg, France.
 \end{acknowledgements}

\bibliographystyle{aa}
\bibliography{mybib.bib}

\end{document}